\documentclass[aps,pra,preprint]{revtex4-1}

\usepackage{amsmath}
\usepackage{amssymb}

\usepackage{xcolor}

\usepackage{setspace}
\usepackage{bm}
\usepackage{graphicx}
\numberwithin{equation}{section}
\usepackage{subfig}

\begin{document}


\title{Engineering Photonic Floquet Hamiltonians through Fabry P\'{e}rot Resonators}
\author{Ariel Sommer \& Jonathan Simon}
\affiliation{James Franck Institute and the Department of Physics at the University of Chicago}
\date{\today}

\begin{abstract}
In this letter we analyze an optical Fabry-P\'{e}rot resonator as a time-periodic driving of the (2D) optical field repeatedly traversing the resonator, uncovering that resonator twist produces a synthetic magnetic field applied to the light within the resonator, while mirror aberrations produce relativistic dynamics, anharmonic trapping, and spacetime curvature. We develop a Floquet formalism to compute the effective Hamiltonian for the 2D field, generalizing the idea that the intra-cavity optical field corresponds to an ensemble of non-interacting, massive, harmonically trapped particles. This work illuminates the extraordinary potential of optical resonators for exploring the physics of quantum fluids in gauge fields and exotic space-times.
\end{abstract}

\maketitle

Time-periodic modulation is under active development both theoretically and experimentally as a tool for Hamiltonian engineering in platforms ranging from cold atoms in optical lattices \cite{GreinerPhotTunnel2011,LewensteinShakenTI2012,BlochReview2012,KolovskyMag2011,HolthausPAT2005,NagerlBloch2008,NagerlSuperBloch2010,TinoSrDeloc2008,KetterleHarper2013} to microwave photons in arrays of superconducting resonators \cite{CarusottoPhotonFluidRMP2013,GirvinCQEDMag2010} and electrons in solids \cite{RefaelTopolaritons2015,GedikFloquetTI2013}. By imposing external fields which couple states of different energies and symmetries, modulation enables time-reversal symmetry breaking and the introduction of synthetic gauge fields \cite{KitagawaFloquet2010,RefaelFloquetTI2011}, as well as manipulation of interactions \cite{SimonThreeBody2014}.

In parallel, there is an aggressive effort to explore optical modes coupled to matter as a platform for quantum manybody phenomenology. Single- \cite{TilmanDickeTransition2010,RitschCavityPhases2007,HelmutQuantizedFields2008,EsslingerDickeRealtime2013} and multi- \cite{LevEmergentCrystal2009,DomokosMultimode2011} mode optical resonators, as well as photonic crystal structures \cite{KimbleAtLightXtal2014,KimbleSubLattice2015} are under investigation to induce long-range interactions between atoms; near-planar resonator/exciton heterostructures \cite{YamamotoRMP2010,YamamotoExcCond2002,KasprzakExcitonBEC2006} and quantum fluids \cite{WeitzPhotonBEC2010} have been employed to study interacting quantum fluids \cite{YamamotoRMP2010,YamamotoExcCond2002,KasprzakExcitonBEC2006}; arrays of microwave resonators coupled to superconducting qubits \cite{HouckChipLatticeReview2012} have been harnessed as a model Bose-Hubbard system; and Rydberg Electromagnetically Induced Transparency (rEIT) in trapped atomic gases has recently been demonstrated as a platform for studying 1D quantum dynamics of strongly interacting photons \cite{VuleticRydbergInt2012,VuleticAttRydPhot2013,LukinRydbergPhotons2011,FleischhauerPhotonCrystal2013,BuchlerScatRes2014}.

Here we formalize a new approach to photonic quantum materials based upon exotic optical resonators; following up on our prior work describing Rydberg-dressed photons in a near-degenerate optical resonator as interacting, massive, harmonically trapped 2D particles in synthetic magnetic fields \cite{SimonCavityPol2015}, we now provide a more detailed framework for designing the resonators and characterizing the resulting single-particle photonic Hamiltonian dynamics.

\begin{figure}
\includegraphics[width=100mm]{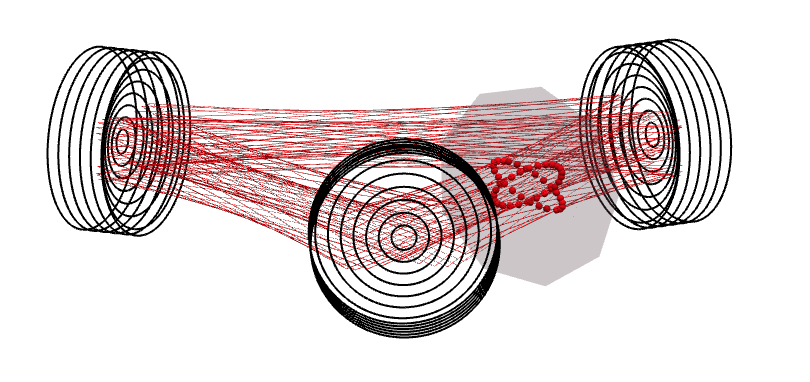}
\caption{\label{SetupFigure} \textbf{Schematic Three-Mirror Fabry-P\'erot Resonator}. A single ray (thin red line) is followed over many-roundtrips through the resonator. The intersection pattern of this ray (red spheres) in a chosen transverse plane (gray polygon) of the resonator traces out a stroboscopic evolution corresponding, in this case, to a massive, harmonically trapped particle, and its image reflected across the origin. These dynamics may be formally understood using the Floquet formalism described in this work.}
\end{figure}

Our approach begins in the ray-optics picture where, assuming round-trip ray-trajectories are nearly closed, round-trip propagation may be coarse-grained using a Floquet formalism to provide an effective 2D time-continuous Hamiltonian for the photon (\textbf{Section \ref{sec:Floquet}}, and Figure \ref{SetupFigure}). We will quantize this Hamiltonian (\textbf{Section \ref{sec:qFloquet}}), leading to a wave-optics view of the resulting physics and the appearance of longitudinal modes due to the energy periodicity of the Floquet formalism. In \textbf{Section \ref{sec:Decomp}} we will classify all of the terms in this Hamiltonian: an inertial mass tensor, harmonic confinement tensor, and a synthetic magnetic field, as well as gauge (non-physical) degrees of freedom. In \textbf{Section \ref{sec:MultiTrip}} we consider what happens when the coarse-graining breaks down because ray-trajectories are not nearly closed, and explore a way to recover a simple Hamiltonian picture if the trajectories nearly close after multiple round-trips.

To illustrate the techniques developed in the preceding sections, we next consider several different resonator geometries (\textbf{Section \ref{sec:examples}}), focusing in particular on the symmetric two-mirror resonator. We distinguish between mechanical- and canonical- ray momentum, and show that while photons in near-planar cavities exhibit a positive mass, those in near-concentric cavities exhibit a negative mass. We then briefly analyze twisted resonators, which introduce synthetic magnetic fields for photons.

Finally, we explore the impact of mirror aberrations and non-paraxial optics on the photonic Hamiltonian (\textbf{Section \ref{sec:ResPerturbations}}). We show that these corrections provide a route to arbitrary potentials and dispersion relations for resonator photons, along with a path to photonic dynamics on curved spatial manifolds.

\section{\label{sec:Floquet}Floquet Formalism for Rays in Optical Resonators}
A paraxial optical resonator may be characterized by an ABCD \cite{SiegmanBook1986} matrix $\bm{M}$, describing the round trip evolution of all light rays in a given transverse plane of the resonator. In particular, the ray described by  $V \equiv \begin{pmatrix}\bm{x}\\\bm{s}\end{pmatrix}$, where $\bm{x}$ is the (2D) transverse location of the ray, and $\bm{s}$ its slope, becomes $\bm{M}V$ under round-trip propagation. This describes a discrete linear transformation in phase space, and suggests that such stroboscopic dynamics (see Figure \ref{fig:Classical1Pass}) are equivalent to periodically sampled continuous evolution under a quadratic time invariant Hamiltonian.

To develop a Hamiltonian formalism describing the continuous evolution of the ray within a particular transverse plane we must first convert the slope $\bm{s}$ into a momentum $\bm{p}$ which is canonically conjugate to $\bm{x}$. This momentum is $\bm{p}=\hbar k \bm{s}$, with $k\equiv2\pi/\lambda$ and $\lambda$ the optical wavelength. We may thus define a phase-space state-vector $\mu\equiv  \begin{pmatrix}\bm{x}\\\bm{p}\end{pmatrix}$, and a round-trip propagation matrix $\bm{B}=\bm{\beta}\bm{M}\bm{\beta^{-1}}$, with $\bm{\beta}\equiv \begin{pmatrix}\bm{I}_2&&\bm{0}\\\bm{0}&&\hbar k\bm{I}_2\end{pmatrix}$ and $\bm{I}_2$ the 2x2 identity matrix. The same round-trip propagation matrix applies to $\bm{x}$ and $\bm{p}$ as operators in paraxial wave optics ~\cite{NienhuisTwistedModes2007}.

Noting that round-trip propagation requires a time $T_{rt}=L_{rt}/c$, where $L_{rt}$ is the total round-trip length along the resonator axis, and $c$ is the speed of light, we may now view $\bm{B}$ as a stroboscopic time evolution operator: $\mu(t+T_{rt})=\bm{B}\mu(t)$. If $\mu$ is to be described by continuous evolution under a general (time-invariant) quadratic Hamiltonian, 

\begin{equation}\label{eqn:Qdef}
H\equiv\frac{1}{2}\mu^\intercal  \bm{G}^\intercal \bm{Q} \mu
\end{equation}
with symmetric $\bm{G}^\intercal \bm{Q}$ and $ \bm{G}=\begin{pmatrix}0&&\bm{I_2}\\-\bm{I_2}&&0\end{pmatrix}$, then Hamilton's equations imply $\frac{d\mu}{dt}=\bm{Q}\mu$. The same result follows using the canonical commutation relations and Heisenberg equations of motion for $\bm{x}$ and $\bm{p}$ as operators.

%

We may now integrate these equations of motion: $\mu(t+\tau)=\exp(\bm{Q}\tau)\mu(t)$. Using $\tau=T_{rt}$ and solving for $\bm{Q}$, we arrive at $\bm{Q}=\frac{c}{L_{rt}}(\log(\bm{B})-2\pi \bm{I} i l)$, where $l \in \mathbb{Z}$, and $\bm{I}$ is the identity matrix. Substituting for $\bm{Q}$ in (\ref{eqn:Qdef}) yields the effective Floquet Hamiltonian:

\begin{equation}\label{eqn:Hdef}
H = \frac{c}{L_{rt}} \left[\frac{1}{2}\begin{pmatrix} \bm{p}^\intercal && -\bm{x}^\intercal \end{pmatrix} (\log{\bm{\beta M \beta^{-1}}}) \begin{pmatrix} \bm{x} \\ \bm{p} \end{pmatrix}-i\pi l\cdot\sum_{i=1}^{2}[x_i,p_i]\right]
\end{equation}


Here we employ the standard definition of the matrix logarithm as the inverse of the matrix exponential, which is itself defined in terms of its Taylor series. Since $\bm{\beta M \beta^{-1}}$ has only eigenvalues of unit modulus, the logarithm will be purely imaginary and due to the branch cut in its domain will be defined only modulo $2\pi i$. So long as $\bm{x}$ and $\bm{p}$ commute (as they do in the ray-optics limit), this term drops out of the Hamiltonian, leaving:
\begin{equation}\label{eqn:Hray}
H_{classical/ray}= \frac{c}{L_{rt}} \left[\frac{1}{2}\begin{pmatrix} \bm{p}^\intercal && -\bm{x}^\intercal \end{pmatrix} (\log{\bm{\beta M \beta^{-1}}}) \begin{pmatrix} \bm{x} \\ \bm{p} \end{pmatrix}\right]
\end{equation}
\section{\label{sec:qFloquet}Quantum Mechanical Treatment}
\begin{figure}[!tbp]
\centering
\subfloat[]{\includegraphics[width=70mm]{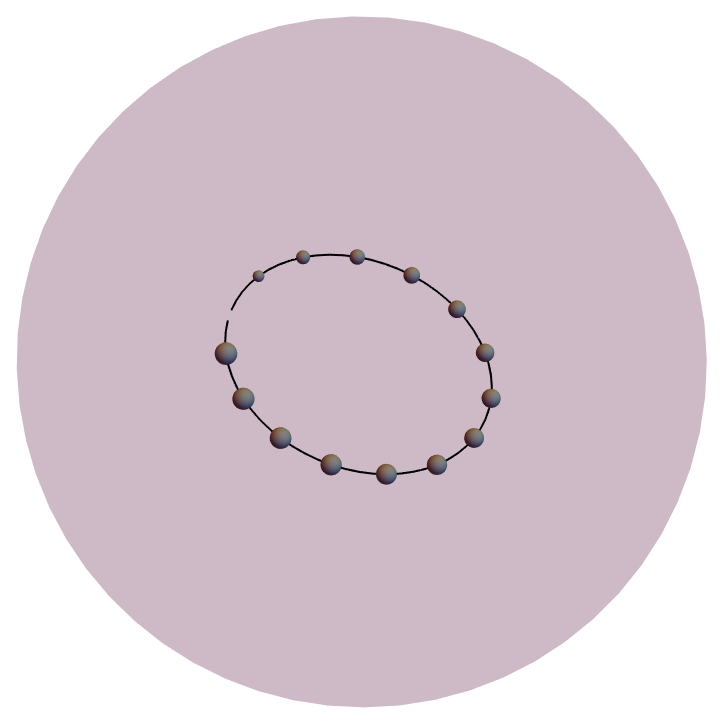}\label{fig:Classical1Pass}}
\subfloat[]{\includegraphics[width=40mm]{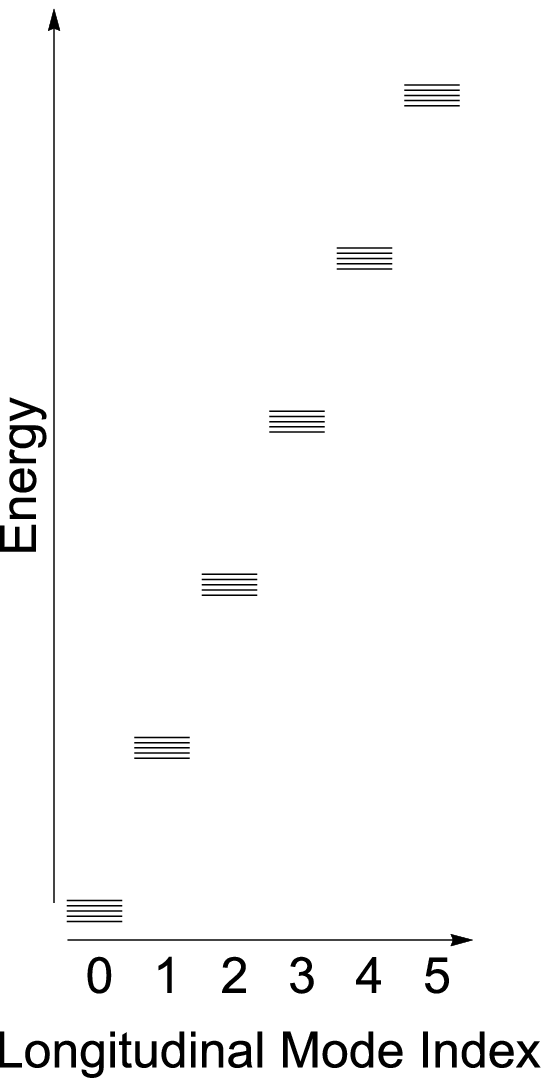}\label{fig:QM1Pass}}
\caption{\textbf{Single-Pass Near-Degenerate Resonators}. \textbf{\protect\subref{fig:Classical1Pass}} Poincar\'e Section in a plane transverse to the propagation axis of a near-planar  near-degenerate optical resonator. The dots indicate discrete transits through the reference plane, with their size reflecting the transit number. The solid curve is the corresponding coarse-grained dynamics in a symmetric harmonic trap. \textbf{\protect\subref{fig:QM1Pass}} Floquet Energy Spectrum for a near-planar, near-degenerate optical resonator. Each near-degenerate manifold is built entirely from states with the same longitudinal quantum number, with a transverse mode-spacing corresponding to the trapping frequency. The energy spacing between Floquet manifolds is given by Planck's constant times the resonator free spectral range, $c/L_{rt}.$}
\label{NearDegenSingleTripFigure}
\end{figure}

Quantizing the Hamiltonian (eqn \ref{eqn:Hdef}) turns the $\bm{x}$'s and $\bm{p}$'s into non-commuting operators. In this case, noting that $[x_i,p_j]=i\hbar\delta_{ij}$, the Hamiltonian becomes:
\begin{equation}\label{eqn:Hwave}
H_{quantum/wave}= \frac{c}{L_{rt}} \left[\frac{1}{2}\begin{pmatrix} \bm{p}^\intercal && -\bm{x}^\intercal \end{pmatrix} (\log{\bm{\beta M \beta^{-1}}}) \begin{pmatrix} \bm{x} \\ \bm{p} \end{pmatrix}\right]+\frac{\hbar c}{L_{rt}}2\pi\cdot l
\end{equation}
The additional $\frac{\hbar c}{L_{rt}}2\pi\cdot l$ in the energy reflects the fact that we are considering a Floquet Hamiltonian \cite{KitagawaFloquet2010,RefaelFloquetTI2011}; the periodic influence of the mirrors on the optical field means that the eigen-frequencies are only defined up to the inverse round-trip time; this is analogous to the quasi-momentum being defined only up to the lattice spacing in a crystal (see Figure \ref{fig:QM1Pass}). It is interesting to note that this energy periodicity corresponds to the cavity free-spectral range, and the $l$ quantum number is actually the familiar longitudinal mode index.
\section{\label{sec:Decomp}Decomposing a General Quadratic Hamiltonian}
A general paraxial Fabry-P\'{e}rot may include non-commuting non-planar reflections and mirror astigmatism, reflected in a near-arbitrary 4x4 ABCD matrix $\bm{M}$, corresponding to a Floquet Hamiltonian $H_{Floquet}= \frac{c}{L_{rt}} \left[\frac{1}{2}\begin{pmatrix} \bm{p}^\intercal && -\bm{x}^\intercal \end{pmatrix} (\log{\bm{\beta M \beta^{-1}}}) \begin{pmatrix} \bm{x} \\ \bm{p} \end{pmatrix}\right]$. We would like to be able to ascertain, for such an arbitrary resonator, what types of dynamics we can engineer, and for a particular resonator, what we \emph{have} engineered.

Because this Hamiltonian must be Hermitian it has only 10 independent parameters, and may be decomposed in the following convenient and physically illuminating way:
\begin{equation}\label{eqn:FullDecomp}
H=\frac{1}{2}(\bm{p}-\beta_k\bm{\sigma}^k\cdot \bm{x})^\intercal\bm{m}_{eff}^{-1}(\bm{p}-\beta_k\bm{\sigma}^k\cdot \bm{x})+
\frac{1}{2}\bm{x}^\intercal\bm{\omega}_{trap}^{\intercal}\bm{m}_{eff}^{-1}\bm{\omega}_{trap}\bm{x}
\end{equation}
with:
$\bm{m}_{eff}^{-1}=\frac{1}{m}\bm{R}_{\theta_m}\cdot\begin{pmatrix}\frac{1}{1+\epsilon_m}&&0\\0&&\frac{1}{1-\epsilon_m}\end{pmatrix}\cdot\bm{R}_{-\theta_m}$,
$\bm{\omega}_{trap}=\omega\cdot\bm{R}_{\theta_t}\cdot\begin{pmatrix}1+\epsilon_t&&0\\0&&1-\epsilon_t\end{pmatrix}\cdot\bm{R}_{-\theta_t}$,
$\bm{\sigma}^k\equiv[\bm{I},\bm{\sigma}^x,\bm{\sigma}^y,\bm{\sigma}^z]$, $\beta_k\equiv[\delta,\Delta_\times,-iB_z/2,\Delta_{+}]$, $\bm{R}_\phi\equiv\begin{pmatrix}\cos{\phi}&&\sin{\phi}\\-\sin{\phi}&&\cos{\phi}\end{pmatrix}$, and the Pauli matrices operate on the real-space vector $\bm{x}$.

The significance and sources of the various terms:
\begin{enumerate}
\item $\bm{m}_{eff}$ and $\bm{\omega}_{trap}$ are the transverse effective mass and trapping frequencies of the particle, respectively, arising from the interplay of cavity length and mirror curvature; they become anisotropic (with axes rotated by $\theta_{m,t}$, and fractional difference $\epsilon_{m,t}$) in the presence of mirror astigmatism, often caused by off-axis reflection \cite{SiegmanBook1986} from otherwise spherical mirrors.
\item The $\beta_k$'s are the remaining 4 degrees of freedom, and parameterize the gauge potential arising from common-mode defocus ($\delta$), rotated differential defocus ($\Delta_{+}, \Delta_\times$), and resonator twist ($B_z$). Defining the vector potential $\bm{A}\equiv\beta_k\bm{\sigma}^k\cdot \bm{x}$, we find $\bm{\nabla}\times\bm{A}=B_z\hat{\bm{z}}$, indicating that the rest of the terms ($\delta, \Delta_+,\Delta_\times$) may be gauged away via $\bm{A}\rightarrow\bm{A}-\bm{\nabla}f$ for $f=\frac{1}{2}\delta(x^2+y^2)+\Delta_\times x y+\frac{1}{2}\Delta_+(x^2-y^2)$. \emph{Thus the only physically significant term is $B_z$, the magnetic field induced by twist}.
\end{enumerate}

The recipe for going from an arbitrary resonator geometry to the physical parameters of the space in which the trapped photons live is to: (1) compute a round-trip 4x4 ABCD matrix for the resonator geometry under consideration, (2) from this compute a Floquet Hamiltonian, and finally (3) decompose this Hamiltonian into the physically significant parameters.

\section{\label{sec:MultiTrip}Near-Degeneracy After Multiple Round-Trips}

The stroboscopic time evolution under the round-trip ray matrix becomes indistinguishable from continuous time evolution in the limit where the transverse dynamics are slow compared to the longitudinal dynamics. In this limit, the frequency splittings between the quantized transverse modes described by the first term in Eqn. (\ref{eqn:Hwave}) become small compared to the splitting between the longitudinal modes described by the second term in (\ref{eqn:Hwave}), and the cavity is said to be nearly degenerate. In such a cavity, rays return to near their original location after each round trip, and the first term in (\ref{eqn:Hwave}) describes the slow precession of the rays after many round trips.

\begin{figure}
\centering
\subfloat[]{\includegraphics[width=70mm]{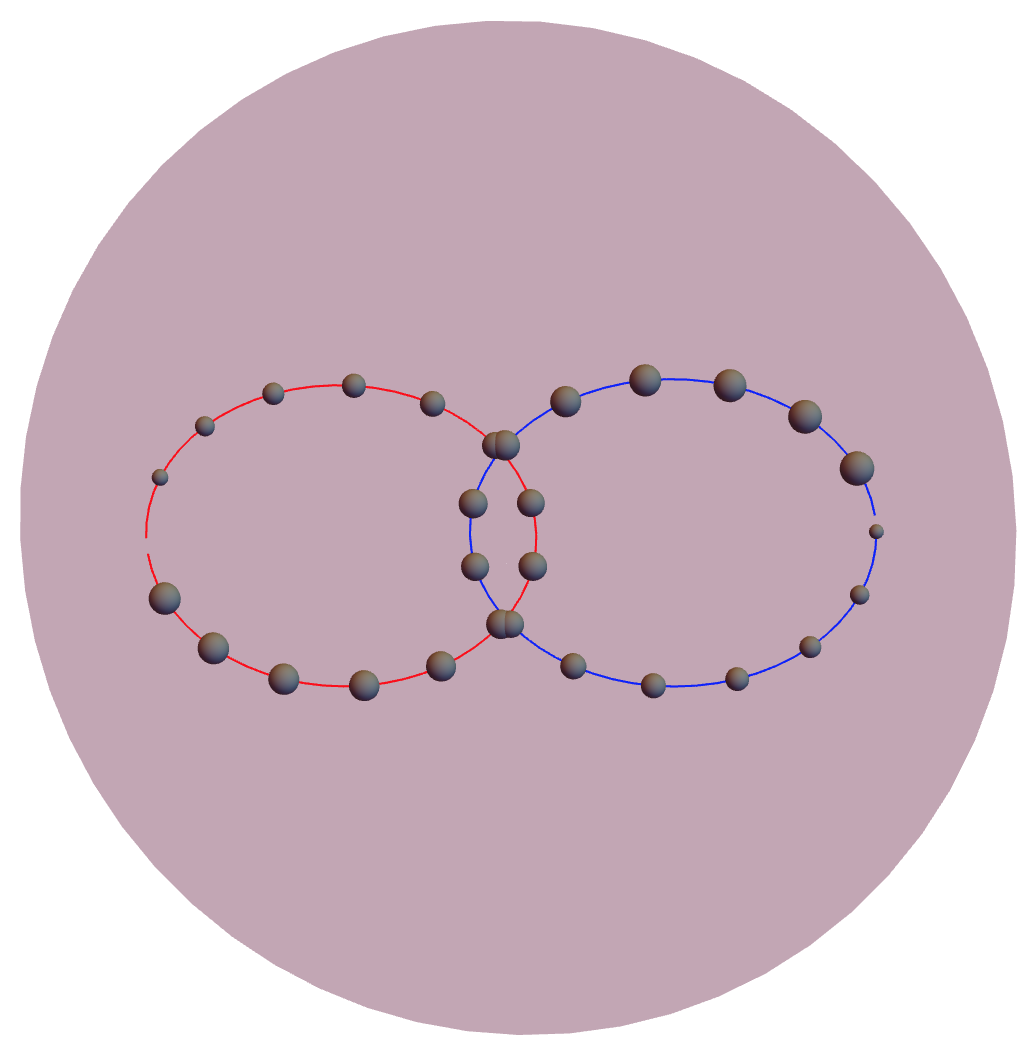}\label{fig:Classical2Pass}}
\subfloat[]{\includegraphics[width=40mm]{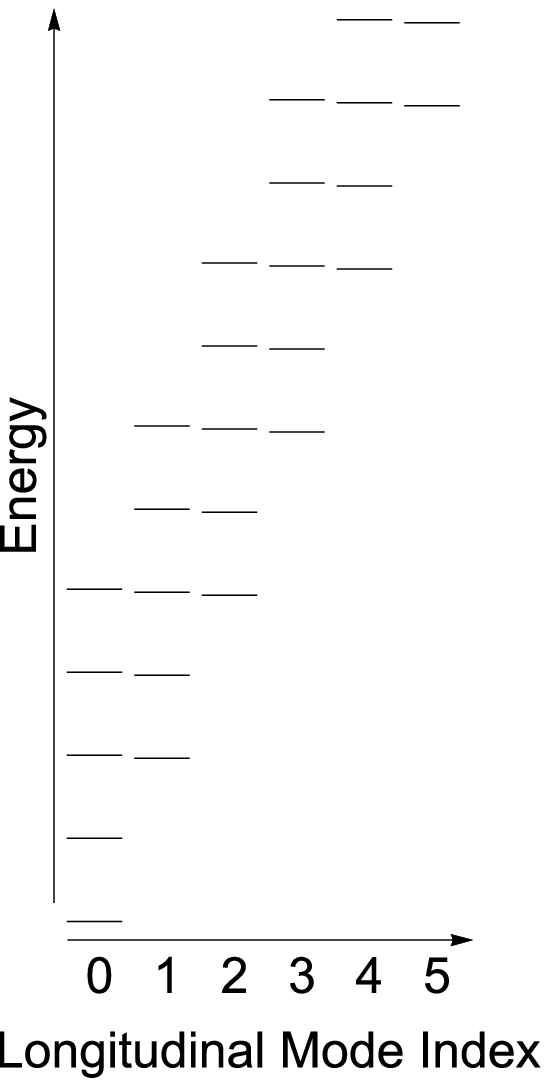}\label{fig:QM2Pass}}
\caption{\label{NearDegenMultiTripFigure}\textbf{Multi-Pass Near-Degenerate Resonators}. \textbf{\protect\subref{fig:Classical2Pass}} Poincar\'e Section in a plane transverse to the propagation axis of a near-confocal near-degenerate optical resonator. The dots indicate discrete transits through the reference plane, with their (decreasing) size reflecting the transit number. The red and blue solid curves reflect the corresponding coarse-grained dynamics for even- and odd- transit numbers. \textbf{\protect\subref{fig:QM2Pass}} Floquet Energy Spectrum for a near-confocal, near-degenerate optical resonator. Each near-degenerate manifold is built from states with many different longitudinal quantum numbers. The energy spacing between Floquet manifolds is given by \emph{half} of the resonator free spectral range.}
\end{figure}

It is often the case that while the ray does not pass near its original phase-space location after a single round-trip, it may do so after several round trips (see Figure \ref{fig:Classical2Pass}). In the wave-optics picture, such a resonator exhibits a Floquet spectrum where near-degeneracy arises from incrementing both longitudinal and transverse quantum numbers in the appropriate proportion (see Figure \ref{fig:QM2Pass}).

More formally, we can write $\bm{M}_{multi}=\bm{M}^s$ and $L_{multi}=s\cdot L_{rt}$; replacing $\bm{M}\rightarrow\bm{M}_{multi}$, $L_{rt}\rightarrow L_{multi}$ in the equations from the preceding sections provides the resulting effective Hamiltonian, with the caveat that the ray appears in every plane with multiple images mirroring its dynamics.

Some typical examples of this phenomenon are (1) the near-confocal resonator, which provides the low energy dynamics of a massive particle in a harmonic trap after two round-trips ($s$=2), and (2) the astigmatism-compensated twisted resonator, which provides the low energy dynamics of a massive particle in a magnetic field after a twist-controllable number of round trips. In the former case, the image rays are reflections across the resonator's longitudinal axis. In the latter case, the image rays are rotated about the resonator's axis.

\section{\label{sec:examples}Examples of Simple Resonators} 
\subsection{Symmetric Two-Mirror Fabry-P\'{e}rot in the Focal Plane}
Here we consider a two-mirror symmetric Fabry-P\'{e}rot resonator, with length L, and mirror radii of curvature R. The round-trip ABCD matrix for a single transverse direction, in the central plane of the cavity (a distance L/2 from each mirror) is (defining $g\equiv1-L/R$):
\begin{equation}
\bm{M}_{focal}=
\begin{pmatrix}
-1+2g^2 && g R (1-g^2)\\
-\frac{4g}{R} && -1+2g^2
\end{pmatrix}
\end{equation}

The matrix logarithm may be obtained using a similarity transform to a rotation matrix, giving
(for $-1\leq g\leq 1$ as required for resonator stability):
\begin{equation}
\bm{Q}_{focal}=\frac{c}{2L}
\begin{pmatrix}
0&&\frac{R\alpha}{2\hbar k}\theta\\
-\frac{2\hbar k}{R\alpha}\theta && 0
\end{pmatrix}
\end{equation}
%
%
with $\alpha\equiv\sqrt{1-g^2}$, $\theta\equiv\cos^{-1}(-1+2g^2)$.
Thus the Floquet Hamiltonian is given by:
$H=\frac{p^2}{2m}+\frac{1}{2}m\omega^2x^2$, where $\omega\equiv\frac{c}{2L}\theta$, and $m\equiv\frac{2\hbar k}{\frac{c}{2L}\alpha R \theta}$.

Quantizing and diagonalizing this Hamiltonian gives a quantum harmonic oscillator $H=\hbar \omega (a^\dagger a+1/2)$ with energy-level spacing $\hbar\omega$, harmonic oscillator lengths $x_0=\sqrt{\frac{\hbar}{m\omega}}=\sqrt{\frac{R\lambda}{4\pi}\sqrt{1-g^2}}$, $p_0=\sqrt{m\hbar\omega}=\sqrt{\frac{4\pi\hbar^2}{\alpha\lambda R}}$, raising operator $a^\dagger\equiv\frac{1}{\sqrt{2}}(\frac{\hat{x}}{x_0}+i\frac{\hat{p}}{p_0})$ and Hermite-Gauss eigenstates $\psi_n=\frac{1}{\sqrt{\sqrt{\pi}2^{n}n!x_0}}e^{-\frac{x^2}{2x_0^2}}H_n(\frac{x}{x_0})$.

These results are consistent with the standard expressions for the two-mirror Fabry-P\'{e}rot \cite{SiegmanBook1986}, where the transverse mode spacing is $\hbar\omega$, and the $1/e^2$ waist of the lowest mode is $w_0=\sqrt{\frac{R\lambda}{2\pi}\sqrt{1-g^2}}=x_0\sqrt{2}$; the factor of $\sqrt{2}$ arises from the different conventions for optical mode waist and harmonic oscillator length.

\subsection{Symmetric Two-Mirror Fabry-P\'{e}rot out of the focal plane}
One might be inclined to ask about the impact upon the transverse Hamiltonian of considering a plane other than the focal plane of the resonator. We will work this out \emph{backwards} first, using knowledge of the resonator eigenmodes and scalar diffraction theory, and then applying the full machinery of the Floquet formalism.

Clearly the eigen-energies of the resonator cannot change (since the eigenstates of the paraxial wave equation are solutions over the full 3D resonator). Furthermore, we know from scalar diffraction theory \cite{SiegmanBook1986} that the impact of diffraction on the mode-functions is (1) a radial rescaling according to $w(z)=w_0\sqrt{1+(\frac{z}{z_r})^2}$; (2) a quadratic wavefront curvature of the form $e^{-i\frac{k x^2}{2\mathcal{R}(z)}}$, for $\mathcal{R}(z)\equiv z \left[1+(\frac{z_r}{z})^2\right]$; and (3) a mode-dependent Gouy phase shift $\zeta_n(z)\equiv n\cdot\tan^{-1}\frac{z}{z_r}$. Here the Rayleigh range is defined by $z_r\equiv\frac{\pi w_0^2}{\lambda}$.

The Gouy phase may be gauged away through a trivial pre-factor on the wavefunction, and we are left with a Hamiltonian system with uniformly spaced eigenvalues and Hermite-Gauss eigenfunctions $\tilde{\psi}_n=\frac{1}{\sqrt{\sqrt{\pi}2^{n-1/2}n!w(z)}}e^{(\frac{-1}{w(z)^2}-\frac{i k}{2\mathcal{R}(z)})x^2}H_n(\frac{x\sqrt{2}}{w(z)})$. 

The question, then, is what Hamiltonian has these mode-functions? We would know (a quantum-harmonic oscillator Hamiltonian $H_{QHO}=\frac{p^2}{2m}+\frac{1}{2}m\omega^2 x^2$, were it not for the wave-front curvature term. We can remove this term by a unitary transformation $U=e^{i\frac{k x^2}{2 \mathcal{R}(z)}}$. The resulting Hamiltonian is $H=UH_{QHO}U^{\dagger}=\frac{(p+\hbar k x/\mathcal{R}(z))^2}{2m}+\frac{1}{2}m\omega^2 x^2$.

We now work forwards, arriving at this Hamiltonian via Floquet techniques. The round-trip ray-matrix for the same two-mirror symmetric Fabry-P\'erot resonator considered previously, but for a plane located at $z=\frac{\epsilon}{2}L$ from the resonator focal plane, is:

\begin{equation}
\bm{M}_{nonfocal}=
\begin{pmatrix}
-1+2g(g-\epsilon(1-g)) && g R (1-g^2+(1-g)^2\epsilon^2)\\
-\frac{4g}{R} && -1+2g^2+2g\epsilon(1-g)
\end{pmatrix}
\end{equation}

A bit of arithmetic yields:
.
\begin{equation}
\bm{Q}_{nonfocal}=\bm{Q}_{focal}+\frac{c}{2L}\epsilon\theta\sqrt{\frac{1-g}{1+g}}
\begin{pmatrix}
1&&(1-g)\epsilon\frac{R}{2\hbar k}\\
0&&-1
\end{pmatrix}
\end{equation}

The off-diagonal correction term in $\bm{Q}_{nonfocal}$ modifies the photon mass, as it impacts only $\frac{\partial^2 H}{\partial p^2}$; it corresponds to a change in the mode-waist due to diffraction. More interesting are the diagonal corrections to $\bm{Q}_{nonfocal}$: $\bm{Q}_{focal}$ lacks any such terms, which correspond to a  $\frac{\partial^2 H}{\partial p \partial x}$ term and reflect a term in the Hamiltonian proportional to $xp$. Following the calculation through, we arrive at:
$H=\frac{(p+b x)^2}{2m}+\frac{1}{2}m\omega^2 x^2$, where $\omega=\frac{c}{2L}\theta$,
$m=\frac{\hbar k}{\frac{c}{4L}R\theta\alpha}\left[\frac{1}{1+\frac{1-g}{1+g}\epsilon^2}\right]$,
$b=\frac{2\hbar k \epsilon}{(1+g)R+(1-g)R \epsilon^2}=\frac{\hbar k}{\mathcal{R}(z)}$ for $\mathcal{R}(z)$ defined as above. This expression coincides with our expectation from the paraxial wave equation, and indeed, the trap frequency does not depend upon defocus.

In short: outside of the focal plane of the resonator, the canonical momentum of the ray (corresponding to its slope as it propagates along the cavity axis) is no longer proportional to the mechanical momentum of the ray (the rate at which it moves in the 2D transverse plane under consideration). Instead, there is an additive correction which is linear in the position, reflecting the wave-front curvature.


\subsection{Near-Concentric vs Near-Planar Fabry-P\'{e}rot}
A separation of timescales between the cavity round-trip time and the harmonic oscillator period requires tuning the cavity geometry near a degeneracy point, where at least one of the transverse mode frequencies becomes much smaller than the cavity free spectral range. Both near-planar ($L\ll R$) and near-concentric ($L\approx 2R$) cavities exhibit such a near-degeneracy. One must be cautious in defining ``near-degenerate,'' however, because while the \emph{ratio} of the transverse- to longitudinal- mode spacing goes to zero in both cases, the transverse spacing \emph{itself} only approaches zero when the appropriate parameter is tuned: the mirror radius of curvature in the near-planar case, and the cavity-length in the near-concentric case.

The trap frequency and mass in the near-planar cavity are: $\omega_{trap}\approx\frac{c}{\sqrt{LR/2}}$, $m\approx\frac{\hbar k}{c}$ . In the near-concentric case they are: $\omega_{trap}\approx\frac{2c}{L}\sqrt{1-\frac{L}{2R}}$, $m\approx -\frac{\frac{\hbar k}{c}}{1-L/2R}$.

Note that:
\begin{itemize}
\item In the near-planar case the mass remains finite, and the trap frequency approaches zero only if ``planarity'' is approached by increasing mirror radius of curvature to infinity rather than by reducing cavity length zero;
\item In the near-concentric case the trap frequency goes to zero and the mass diverges, no matter how one approaches degeneracy (by adjusting resonator length, or mirror curvature);
\item The photon mass is \emph{negative} in the near-concentric case. This reflects the fact that the direction of ray propagation out of the plane is opposite to the direction of motion of the particle within the plane (canonical and mechanical momenta are opposite), due to an inversion from the refocusing of the cavity mirrors.
\end{itemize}

\subsection{Twisted Resonators}
\begin{figure}
\includegraphics[width=100mm]{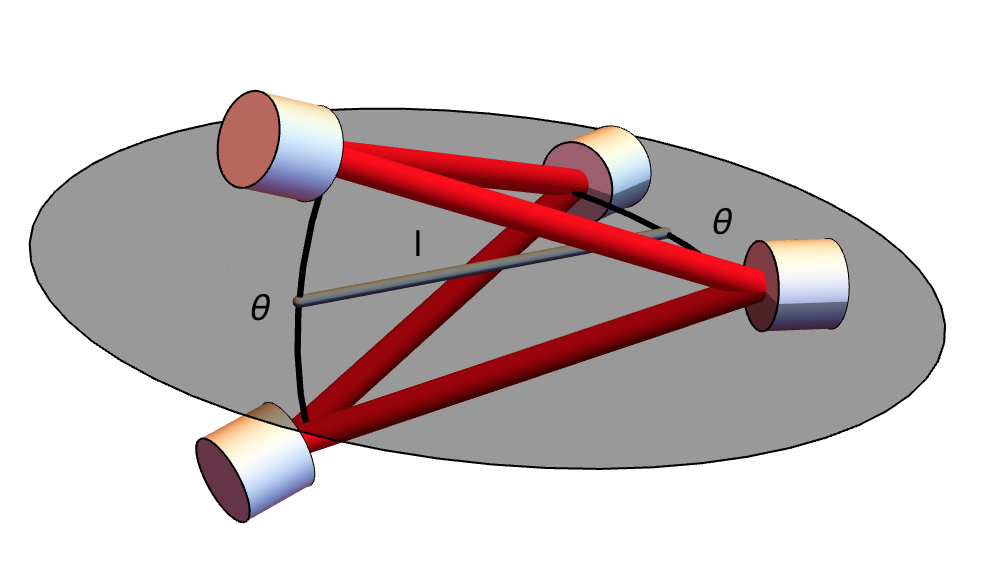}
\caption{\label{TwistedCavityFigure} \textbf{Schematic non-planar resonator}. $\theta$ is the cavity opening angle, and $l$ is the on-axis length. The out-of-plane reflections result in an image rotation, akin to a pair of dove-prisms, on each round-trip through the resonator. This rotation is equivalent to Coriolis and centrifugal couplings in the Floquet Hamiltonian, the former of which may in turn be interpreted as a uniform synthetic magnetic field for the cavity photons.}
\end{figure}

The simplest resonators that exhibit synthetic magnetic fields for the photons traveling within them are (a) four-mirror resonators that do not reside in a plane, and (b) three-mirror resonators with astigmatic mirrors that are twisted with respect to one another. What these resonator geometries have in common is a helicity to the round-trip manipulation of the photon trajectory, producing dynamics akin to a Floquet topological insulator \cite{KitagawaFloquet2010}. Because (a) is easier to realize experimentally, it is the path that we will explore here (see Figure \ref{TwistedCavityFigure}).

We consider a four-mirror resonator where the mirrors do not all reside in a plane, but where all mirrors are curvature-less (``planar''), to keep the analysis simple. As shown in the figure at right, the resonator geometry consists of an opening angle $\theta$, and a principal arm length $l$. To analyze such a resonator requires 4x4 ABCD matrices and careful transformation of coordinate bases at each reflection. The outcome is:
$\bm{M}=\begin{pmatrix}1&&l\\0&&1\end{pmatrix}\otimes \bm{R}(\phi)$, where $\bm{R}(\phi)$ is a 2D rotation through an angle $\phi$ given by: $\cos{\phi}\equiv\frac{1}{16}(3+8\cos{\theta}+12\cos{2\theta}-8\cos{3\theta}+\cos{4\theta})\approx \theta^2$ for small $\theta$.

It is thus apparent that the resonator rotates the coordinate axes on each round-trip; the Floquet Hamiltonian is:
\begin{equation}
\label{eqn:HTwist}
\begin{split}
H_{Floquet}&=\frac{1}{\sqrt{1+2\tan^2{\frac{\phi}{2}}}}\left[\frac{1}{2}\frac{c}{\hbar k}(p_x^2+p_y^2)+\frac{c}{2 l} \phi (y p_x- x p_y)\right]=\left[\frac{p^2}{2 m_{eff}}-\frac{q B_{eff}}{2 m_{eff}}L_z\right]\\
&=\frac{(\bm{p}-\frac{1}{2}q B_{eff}\hat{\bm{z}}\times\bm{r})^2}{2m_{eff}}-\frac{1}{2}m_{eff}\omega_{trap}^2 r^2
\end{split}
\end{equation}
with an effective mass of $m_{eff}=\frac{\hbar k}{c}\sqrt{1+2\tan^2{\frac{\phi}{2}}}$, and rotation giving rise to an effective magnetic field $q B_{eff}=\frac{\hbar k}{l}\phi$ and corresponding centrifugal anti-trapping with frequency $\omega_{trap}=\frac{1}{\sqrt{1+2\tan^2\frac{\phi}{2}}}\frac{c}{2 l}\phi$. The magnetic length corresponding to the synthetic field is $l_B\equiv\sqrt{\frac{\hbar}{\left|qB\right|}}=\sqrt{\frac{l \lambda}{2 \pi \phi}}$. Note that the effective photon charge $q$ and synthetic magnetic field $B_{eff}$ are individually meaningless (and arbitrary); only the product $q B_{eff}$ is well defined.

In practice mirror curvature is essential to compensate the centrifugal anti-trapping; analysis of curved-mirror non-planar resonators in the presence of non-normal-reflection-induced astigmatism is beyond the scope of this work, and will be presented in a separate publication. 
\section{\label{sec:ResPerturbations}Higher-Order Perturbations to the Resonator} 
Thus far we have analyzed the Hamiltonian that results from light trapped within a resonator composed entirely of quadratic optics, with paraxial (quadratic) propagation between these optics. It is well-known that resonator mirrors are measurably imperfect \cite{KlaassenFP2006}, both because they are spherical rather than parabolic and due to wavefront errors. Furthermore, the propagation of optical fields only approximately obeys the paraxial wave equation, with corrections arising at the same order as spherical aberration on the mirrors $\phi_{guoy}\propto\frac{\lambda}{R_{mirror}}$. Such corrections already become relevant even for low-order modes in moderate finesse $F\sim10^4$ resonators (for 1cm ROC mirrors), and become increasingly important for higher order modes. There are a several calculations of the resonator modes in the presence of such corrections \cite{ZeppenfeldFPCavity2010,KlaassenNonParaxial2004}; here we instead compute the impact of such higher-order terms on the Floquet Hamiltonian.

\subsection{Impact of Non-Quadratic Optics on Trapping Potential, Single-Particle Dispersion, and Spatial Curvature}
Consider an arbitrary lens providing a position dependent phase shift $\alpha(x)$ in a transverse plane that is a longitudinal distance $z$ from the plane where the Floquet Hamiltonian is defined. If the lens is weak enough that it couples only within Floquet bands, but not between them, its impact may be written as a perturbation to the Floquet Hamiltonian itself. In fact, we rely upon such corrections to truncate the Floquet energy spectrum before the Floquet bands overlap.

The simplest case is $z=0$; a lens in the plane where the Floquet Hamiltonian is defined (henceforth the ``Floquet plane''). In this situation the phase shift $\alpha(x)$ per round trip corresponds to an energy (and thus effective potential) $\frac{\hbar c}{L_{rt}}\alpha(\hat{x})$; the lens directly acts as a potential for the cavity photons. Note that $\alpha(\hat{x})$ is evaluated using the Taylor series expansion of $\alpha(x)$.

If the lens were placed within the cavity in a Fourier plane, the phase shift would be dependent upon the \emph{momentum} in the Floquet plane, and the corresponding Hamiltonian term would be $\frac{\hbar c}{L_{rt}}\alpha(\frac{f}{\hbar k}\hat{p})$, where $f$ is effective focal length of the real-to-Fourier space imaging. Thus in this case, the lens acts to control the \emph{dispersion} of the photons in the Floquet plane.\\\\
It is natural to ask what happens if the lens is placed between real- and Fourier- planes; We will find, in accordance with our ray-optics expectation, that the correction is $\frac{\hbar c}{L_{rt}}\alpha(\hat{x}+\frac{z}{\hbar k}\hat{p})$ 
We now perform the simplest version of the aforementioned calculation: for an arbitrary lens a distance $z$ from our Floquet plane, in one transverse dimension. More sophisticated calculations in two transverse dimensions with an arbitrary ABCD matrix in-between, are simply extensions of this technique.\\\\
Consider the Hamiltonian for an arbitrary lens in the plane z, which produces a round-trip phase-shift of $\alpha(x)$. We can compute its expansion in the plane at $z=0$ by inserting identity operators:
\begin{equation}
\begin{split}
H_{lens}&=\frac{\hbar c}{L_{rt}}\alpha(\hat{x};z)=\frac{\hbar c}{L_{rt}}\int \! \alpha(x) \left|x;z\right\rangle\left\langle x;z\right| \, \mathrm{d}x\\
&=\frac{\hbar c}{L_{rt}}\int \! \alpha(x) \left|x_1;z=0\right\rangle\left\langle x_1;z=0 | x;z\right\rangle\left\langle x;z | x_2;z=0\right\rangle\left\langle x_2;z=0\right| \, \mathrm{d}x \,\mathrm{d}x_1 \, \mathrm{d}x_2
\end{split}
\end{equation}
We now relate localized excitations in the different planes via the free-space Green-function in the paraxial (Fresnel) approximation \cite{AgarwalFresnel1996,GoodmanFourierOptics1969}: $\left\langle x_1;z=0 | x;z \right \rangle=\frac{e^{i k z}}{\sqrt{i\lambda z}}e^{\frac{i k }{2 z}(x-x_1)^2}$.
\begin{equation}
H_{lens}=\frac{\hbar c}{L_{rt}}\frac{1}{\lambda z}\int \! \alpha(x) e^{\frac{i k }{2 z}(x-x_1)^2}e^{\frac{-i k }{2 z}(x-x_2)^2} \left | x_2;z=0\right\rangle\left\langle x_2;z=0\right| \, \mathrm{d}x \,\mathrm{d}x_1 \, \mathrm{d}x_2
\end{equation}
redefining $x_j\rightarrow x_j+x$ we have:
\begin{equation}
H_{lens}=\frac{\hbar c}{L_{rt}}\frac{1}{\lambda z}\int \mathrm{d}x_1 \, \mathrm{d}x_2 e^{\frac{i k }{2 z}(x_1^2-x_2^2)}\int\mathrm{d}x\alpha(x) \left | x_1+x;z=0\right\rangle\left\langle x_2+x;z=0\right|
\end{equation}
and identifying: 
\begin{equation}
\int\mathrm{d}x\left| x_1+x;z=0\right\rangle\alpha(x)\left\langle x_2+x;z=0\right|=e^{i\hat{p} x_1}\alpha(\hat{x})e^{-i\hat{p}x_2}=e^{i \hat{p}(x_1-x_2)}\alpha(\hat{x}+x_2)
\end{equation}
yields:
\begin{equation}
H_{lens}=\frac{\hbar c}{L_{rt}}\frac{1}{\lambda z}\int \mathrm{d}x_1\mathrm{d}x_2 e^{\frac{i k}{2 z}(x_1^2-x_2^2)}e^{i\hat{p}(x_1-x_2)}\alpha(\hat{x}+x_2)
\end{equation}
Performing the $x_1$  integral yields:
\begin{equation}
\begin{split}
H_{lens}&=\frac{\hbar c}{L_{rt}}\frac{\sqrt{i\lambda z}}{\lambda z}\int\mathrm{d}x_2 e^{-\frac{i k}{2 z}(x_2-\frac{z}{\hbar k}\hat{p})^2}\alpha(\hat{x}+x_2)\\
&=\frac{\hbar c}{L_{rt}}\frac{1}{\sqrt{-i\lambda z}}\int\mathrm{d}x_2 e^{-\frac{i k}{2 z}(x_2-\frac{z}{\hbar k}\hat{p})^2}\alpha(\hat{x}+x_2)e^{\frac{i k}{2 z}(x_2-\frac{z}{\hbar k}\hat{p})^2}e^{-\frac{i k}{2 z}(x_2-\frac{z}{\hbar k}\hat{p})^2} \\
&=\frac{\hbar c}{L_{rt}}\frac{1}{\sqrt{-i\lambda z}}\int\mathrm{d}x_2 \alpha\left(e^{-\frac{i k}{2 z}(x_2-\frac{z}{\hbar k}\hat{p})^2}\hat{x}e^{\frac{i k}{2 z}(x_2-\frac{z}{\hbar k}\hat{p})^2}+x_2\right)e^{-\frac{i k}{2 z}(x_2-\frac{z}{\hbar k}\hat{p})^2}\\
&=\frac{\hbar c}{L_{rt}}\frac{1}{\sqrt{-i \lambda z}}\alpha(\hat{x}+\frac{z}{\hbar k}\hat{p})\int \mathrm{d}x_2 e^{-\frac{i k}{2 z}(x_2-\frac{z}{\hbar k}\hat{p})^2}\\
&=\frac{\hbar c}{L_{rt}}\alpha(\hat{x}+\frac{z}{\hbar k}\hat{p})
\end{split}
\end{equation}

Where we have (1) inserted an identity operator; (2) required that $\alpha(x)$ be analytic; (3) used the Baker-Campbell-Hausdorf formula; and (4) performed the remaining Gaussian integration. In two transverse dimensions it may be shown that an arbitrary lens produces a correction to the Hamiltonian: $\frac{\hbar c}{L_{rt}}\alpha(\bm{\hat{x}}+\frac{\bm{\hat{p}}}{\hbar k})$.

We now consider the simple case of two quartic lenses $\alpha(x)=\beta x^4$ placed symmetrically around z=0: $\alpha(\bm{x}-\mu\bm{p})+\alpha(\bm{x}+\mu\bm{p})$. The resulting Hamiltonian contains terms quartic in $\bm{x}$, which we view as a quartic confining potential, terms quartic in $\bm{p}$, which we view as quartic single-particle dispersion, and those quadratic in both $\bm{x}$ and $\bm{p}$ corresponding to manifold curvature. Matching terms in the (non-relativistic) geodesic equation yields a scalar curvature $R=\frac{2r^2/r_0^2}{\left(1+(\frac{r}{r_0})^2\right)^2\left(1+3(\frac{r}{r_0})^2\right)^2}$, where $r_0^2\equiv\frac{1}{\beta\mu^2}$, from a metric:

\begin{equation}
\mathrm{d}s^2=(\mathrm{d}x^2+\mathrm{d}y^2)\frac{x^2+y^2}{r_0^2}+\frac{2(x\,\mathrm{d}x+y\,\mathrm{d}y)^2}{r_0^2}-c^2\mathrm{d}t^2
\end{equation}

Note that an arbitrary perturbation in a \emph{single} plane may always be understood (through a linear canonical transformation) as a real-space potential- two such perturbations (in different planes) are required to generate curvature that cannot be trivially removed through such a generalized coordinate transformation.

One might be inclined to attempt to draw a parallel to a Trotterized cold atom implementation, where atoms are allowed to evolve in a harmonic trap, and then a quarter of a trap period later, when they are in momentum space, an optical potential is briefly applied to them with the hope that it would provide a effective momentum-dependent force (when viewed another quarter trap cycle later). A simple calculation reveals that this does not work, because the atoms evolve back to real space in a way that depends upon the optical potential applied to them. The key to this idea working for photons where it fails for atoms is that the potential may be weak enough that it does not appreciably impact the photons within a single round-trip (it does not mix different Floquet/longitudinal manifolds), but is still strong enough to substantially change the transverse dynamics within a single near-degenerate Floquet manifold; in short, the cavity photons live \emph{simultaneously} in real space, momentum space, and everywhere in-between.\\
\subsection{Beyond Paraxial Optics}
Taking the scalar wave equation $c^2\nabla^2\psi=\partial_t^2\psi$, and substituting $\psi=\phi e^{i k (z-c t)}$ yields:
$(\nabla_\perp^2+2 i k\partial_z+\partial_z^2)\phi=0$.
At lowest order in $\frac{\partial_z}{k}$ this reduces to the paraxial wave equation: $(\nabla_\perp^2+2 i k \partial_z)\phi=0$.

Iterating this approximate solution for the second $z$ derivative yields a first Born correction: $(2 i k \partial_z+\nabla_\perp^2-\frac{(\nabla_\perp^2)^2}{4 k^2})\phi=0$.

This expression says that, up to a constant offset, the phase acquired from propagation along an arm of the cavity of length $l$, is $k_z l =\frac{1}{2}k l \left[(\frac{p}{\hbar k})^2+\frac{1}{4}(\frac{p}{\hbar k})^4\right]$. The first (quadratic) term gives rise to the kinetic dynamics we have been studying throughout this work, and the second (quartic) term is a new correction. Note that this expression is equivalent to a Taylor expansion of $k_z l = l \sqrt{(\hbar k)^2-p^2}$ to quartic order.

Because the ray momentum is transformed after each mirror reflection, the total non-paraxial correction, arising from the term in each arm of the cavity, is:

\begin{equation}
H_{non-paraxial}=\frac{\hbar c}{L_{rt}}\int\mathrm{d}z\frac{(\nabla_\perp^2)^2}{8 k^3}=\frac{c}{8 (\hbar k)^3}\sum_j\epsilon_j(\bm{D}_j\hat{\bm{p}}+\hbar k \bm{C}_j \hat{\bm{x}})^4
\end{equation}

Here we have employed ABCD matrices that move from the Floquet plane to the region between the $j^{th}$ and $(j+1)^{st}$ mirrors and $\epsilon_j$, the fraction of the path length between those two mirrors.

We thus see that the lowest order correction to paraxial optics introduces a quartic potential, quartic dispersion, and cross terms (including manifold curvature), akin to an out-of-focus quartic lens.
\section{\label{sec:Conclusion}Conclusion}
In this paper, we have harnessed the fact that an optical resonator may be viewed as a periodic drive applied to a 2D optical field to develop a Hamiltonian formalism for understanding photonic dynamics in such resonators. This approach applies both within the paraxial, quadratic approximation, where it results in arbitrary quadratic Hamiltonians tunable through resonator geometry, and to perturbations which extend beyond the paraxial limit and produce exotic photon traps, dispersions, and manifold curvatures. This work points to fascinating studies of wave dynamics on curved manifolds, and in conjunction with Rydberg EIT to induce interactions between photons, an exciting route to strongly correlated photonic quantum materials, including those in the presence of synthetic gauge fields and  manifold curvature.

\section{\label{sec:Acknowledgemends}Acknowledgements}
We would like to thank Brandon Anderson, Michael Levin, and Paul Weigmann for fruitful discussions. This work was supported by AFOSR grant FA9550-13-1-0166, and DARPA grant D13AP00053.

\bibliography{TestBib3}{}

\end{document}